\documentstyle[preprint,aps]{revtex}
\begin{document}
\preprint{NIIG-DP-94-3}

\title{An explanatory study of $^{11}$Li $\beta$ decay into
\\$^9$Li and deuteron}
\author{Y. Ohbayasi and Y. Suzuki$^{a)}$}
\address{Graduate School of Science and Technology,
Niigata University,
         Niigata 950-21, Japan}
\address{$^{a)}$Department of Physics, Niigata University,
        Niigata 950-21, Japan}
\date{November,1994}
\maketitle

\begin{abstract}
The $\beta$-decay of $^{11}$Li into $^9$Li and {\it d} is
studied using simple halo wave functions of $^{11}$Li.
 The sensitivity of the transition probability is elucidated on
a description of the halo part and on a choice of the potential
between $^9$Li and {\it d} .
\end{abstract}

\narrowtext
Experiment with radioactive nuclear beams led to the
discovery\cite{Ta85} of light nuclei with
 unusually large matter radii near the neutron drip-line. The nucleus
$^{11}$Li has received much attention as one of those typical systems
which have neutron halo-like structure. The direct information on the
extended structure was derived from the enhancement of the interaction
cross sections.  Momentum distributions of fragments have offered
another useful way of probing the correlated structure of the halo
part and testing model descriptions.  Beta-delayed particle emission
from the nuclei near the drip-line also provides us with a unique
probe for the halo structure. The $^6$He $\beta$ decay into $\alpha$
and {\it d} with the branching ratio,
P$_d$=(7.6$\pm$0.6)$\cdot$10$^{-6}$,
for $E_d > 350$keV\cite{Bo93} is a
good example and has recently been studied by several
authors\cite{Zh93,Baye94,Csoto94,Bark94}.
 It was stressed in ref.\cite{Baye94}, among others, that this
decay is quenched by a cancellation between the internal and halo
contributions to the Gamow-Teller(GT) matrix element and shows an
extreme sensitivity to the halo description up to large distances.
The purpose of this letter is to do an explanatory analysis
 on the $^{11}$Li $\beta$ decay into $^9$Li and {\it d}.
This decay attracts much attention and is expected
to give important information on
the halo structure of $^{11}$Li.

Most of the $\beta$ decay($\sim$97\%)
of $^{11}$Li takes
place to low-lying states in $^{11}$Be.
A small branch of the $\beta$ decay involves delayed {\it n},
{\it t} and $\alpha$ emission. Because of the experimental difficulty
there is no direct evidence
for the $\beta$-delayed {\it d} emission from $^{11}$Li .
There is, however, only one experiment\cite{Wil91} that has
 measured the {\it t}+({\it d}) spectrum down to 350 keV although
no separation was made
 between {\it t} and {\it d}. The experiment indicates
that two separate components
 appear in the spectrum. The intensity at lower energy may
be attributable to the
 branch of the deuterons because the {\it Q} values
for $^8$Li+{\it t} and $^9$Li+{\it d}
 decay channels are 4.96
 MeV and 2.76 MeV, respectively.
The lower limit on the branching ratio determined
 from the experiment is
P$_{t+(d)}$=(1.8$\pm$ 0.3)$\cdot$10$^{-4}$ and
a tentative value for the deuteron
 branch is obtained as P$_d$=(1.0$\pm$ 0.3)$\cdot$10$^{-4}$.

Despite numerous investigations the understanding of
the structure of $^{11}$Li is still
 imperfect. This is in sharp contrast to the case of
$^6$He in which a $^4$He+{\it n}+{\it n} three-body
 model works nicely. The three-body configuration of
$^9$Li+{\it n}+{\it n} is certainly an important
 ingredient for the $^{11}$Li structure but there are
at least two basic points to be settled before a more
 satisfactory description of $^{11}$Li is made, i.e.,
the role of the polarizability of $^9$Li and the feature
of the $^9$Li-{\it n} interaction.
At present we have to use those very simple wave
functions for $^{11}$Li
which assume a $^9$Li core,
and our study is therefore of qualitative nature.

We assume that the $\beta$-delayed {\it d} emission proceeds through
 the so-called direct decay, i.e., a transition from $^{11}$Li to
$^9$Li+{\it d} continuum. In this $\beta$ decay the
 isospin changes and the total angular momentum,
$J_f$, in the final channel may be $J_f$=1/2,
 3/2, or 5/2. Since the decay {\it Q}-value is small,
we assume the relative motion between $^9$Li
 and {\it d} is an {\it s}-wave and independent of $J_f$.
Under these assumptions the decay transition
 probability per time and energy units, {\it dW/dE},
is calculated by\cite{Baye88,Baye94}

\begin{equation}
\frac{dW}{dE}= \frac{mc^2}{\pi^4 v \hbar^2} G_{\beta}^2
f(Q-E)  B_{GT}(E),
\label{eq.1}
\end{equation}
\noindent
where {\it m} is the electron mass, {\it v}
the relative velocity between
 $^9$Li and {\it d}, $G_\beta$=2.996$\times$10$^{-12}$
the dimensionless $\beta$-decay constant,
and {\it f} is the Fermi integral.
The GT reduced matrix element
is expressed as
\begin{eqnarray}
B_{GT} (E) = 6 \lambda^2 < g_E | \psi_{eff}  >^2,
\label{eq.2} \end{eqnarray}
where $\lambda=-1.25$ is the ratio of the axial-vector
to vector coupling constants.
The radial part of the $^9$Li-{\it d} relative motion function
$g_E$ is normalized as
\begin{equation}
g_E(R)
\raisebox{-1.5ex}{\mbox{$\left. \begin{array}{c}
\rightarrow \\ \raisebox{.5ex}{\mbox{$^{R \rightarrow \infty}$}}
\end{array} \right.$}}R^{-1} (F_0 (kR) \cos \delta + G_0 (kR)
\sin \delta),
\label{eq.3}
\end{equation}
where {\it k} is the wave number of
the relative motion with energy {\it E}
and $\delta$ is the {\it s}-wave phase shift at energy {\it E}.
The effective wave function $\psi_{eff}$ is defined as
\begin{equation}
\psi_{eff}(R) = {\int}_0^{\infty} F_d(r) \chi_0(r,R) r^2 dr,
\end{equation}
where $F_d(r)$ is the radial part of the deuteron
wave function. The function
$\chi_0(r,R)$
is the {\it s}-wave part in both of the coordinates,
$\mbox{\boldmath $r$}$ and $\mbox{\boldmath $R$}$,
of a two-neutron reduced amplitude
$\chi(\mbox{\boldmath $r$},\mbox{\boldmath $R$})$ defined by
\begin{eqnarray}
\chi(\mbox{\boldmath $r$},\mbox{\boldmath $R$} )
=(55)^{1/2}
<\phi_{J=\frac{3}{2}M}(^9{\rm Li}) , SM_S = 00, TM_T = 11
| \Psi_{\frac{3}{2}M}(^{11}{\rm Li})>,
\label{eq.6}
\end{eqnarray}
\noindent
where $\mbox{\boldmath $r$}$ is the relative coordinate
between the two neutrons and
$\mbox{\boldmath $R$}$ is the relative coordinate
between $^9$Li and the center-of-mass of the two neutrons.

We used the $^9$Li-{\it d} relative motion function $g_E$
generated from the folding potential of Watanabe type
\begin{equation}
V(R)=\frac{1}{4\pi}\int {F_d}^2 (r)
\left[U_p (
     |\mbox{\boldmath $R$}+
\frac{\mbox{\boldmath $r$}}{2}|) +
 U_n(|\mbox{\boldmath $R$}
-\frac{\mbox{\boldmath $r$}}{2}|)
\right] d\mbox{\boldmath $r$},
\label{eq.7}
\end{equation}
where $U_p(U_n)$ is a {\it p}({\it n})-$^9$Li
optical potential of Woods-Saxon form. We assumed
$U_p$=$U_n$ and took into account only the real part of
the central potential($V_0$=53 MeV).
Dependence of the transition probability on
the optical potential parameters,
the diffuseness {\it a} and the radius parameter $r_0$, will be
discussed later.
The Coulomb potential of uniform charge distribution was used.
As stated in
the beginning our purpose is not to make a quantitative prediction
but to learn the
sensitivity of the spectrum and the branching ratio on
the halo structure and the continuum wave function. We
consider following simple two-neutron reduced amplitudes
which were employed in the
analysis\cite{Ogawa94} of the momentum distribution of
a $^9$Li fragment arising from the $^{11}$Li reaction:
(i) a (p$_{1/2}$)$^2_{J=0}$
harmonic-oscillator shell model(SM) wave function,
 (ii) a di-neutron cluster model(CM) wave function,
and (iii) a hybrid model(HM) wave function.
For each case the function $\chi_0$ is expressed in terms of
a nodeless harmonic oscillator function,
$F(b,r) = 2 \pi^{-1/4} b^{-3/2} \exp(- r^2/2 b^2)$\cite{Baye94}:
\begin{equation}
\chi^{SM}_0 (r, R) =
\frac{1}{6} b^{-2}_0 (r^2 - 4R^2) F (2^{1/2} b_0,
r) F (2^{-1/2} b_0, R),
\end{equation}
\begin{equation}
\chi^{CM}_0 (r, R) = F (b_1 , r) F (b_2 , R).
\end{equation}

\noindent
The parameters, $b_0$ and ($b_1$,$b_2$), are chosen
to reproduce the measured interaction cross section of $^{11}$Li on
$^{12}$C at an incident energy of 800 MeV/nucleon.
See Table 1 of ref.\cite{Ogawa94}
for the values.
These two-neutron halo wave functions,
however, cannot explain the observed momentum distribution of
$^9$Li in the ($^{11}$Li, $^9$Li)
reaction. In case (iii) a simple wave function that reproduces
the momentum distribution was
constructed\cite{Ogawa94} by taking a linear combination of
the shell model and cluster model wave
functions:
$\chi^{HM}_0(r,R)=
\varepsilon_1 \chi^{SM}_0(r,R) + \varepsilon_2 \chi^{CM}_0(r,R)$
 with $\varepsilon_1=0.79$ and $\varepsilon_2=-0.41$.
The hybrid model is considered "best" among three
cases.

Fig.1 illustrates how the functions $g_E$ and $\psi_{eff}$
contribute to the GT matrix element
 for {\it E}=0.5 MeV ( $a$=0.6 fm, $r_0$=1.2 fm). The integral,
$I(R)={\int}^R_0\:g_E(R')\:\psi_{eff}(R')\:{R'}^2 dR'$, for
three effective wave functions is
displayed in the lower part of the figure.
The limit of {\it I(R)} for $R\rightarrow \infty$
appears in eq.(2).
The GT value gains most of the contribution in the region
of {\it R}=4$\sim$10 fm. It has negligible contribution from
the interior part of the wave function,
which is understood by the fact that the $^9$Li-{\it d}
relative motion function exhibits more
oscillatory behavior in the interaction region
than the $\alpha$-{\it d} case.
We can conclude that the $\beta$ decay of $^{11}$Li
into $^9$Li and {\it d} probes mainly the halo part of the
$^{11}$Li wave function. This is in sharp
contrast to the case of $^6$He, where the cancellation in
the GT matrix element occurred
between the internal and external parts of the $^6$He wave function.

We have compared the energy dependence of the transition
probability for three model
wave functions.
All three cases yielded similar spectrum although the magnitudes
of the transition probability differed within
a factor of ten at the peak values.

Table \ref{Tab.1} lists the branching ratio calculated
with the HM wave function for various values
of the potential parameters, {\it a} and $r_0$.
Since no $^9$Li+{\it d} scattering data are available,
we cannot select a good set of the parameters.
The potential with $r_0$=1.3 fm is ruled out,
however, because it leads to a smaller branching ratio
than expected from experiment.
Increasing $r_0$ or $a$ leads to a stronger potential.
The branching ratio of order 10$^{-6}$ is obtained when the
third node of $|g_E|$ is pulled inside at shorter distance
and the overlap integral with $\psi_{eff}$ becomes smaller
as the potential becomes more attractive.
Fig.2 shows the dependence of the transition probabilities
on the potential parameters.
The energy  at which the peak
of the spectrum occurs is dependent on the choice of the potential.
Measurements with
good statistics down to the lower energy will be very
informative in this respect.

In conclusion we have examined
the $\beta$ decay of $^{11}$Li
into $^9$Li and {\it d} in order to shed light on the role of
the halo structure of $^{11}$Li.
We find that the $\beta$-decay matrix
elements are determined practically
by the halo part of distances, $R\ge4$ fm.
The branching ratio and the $^9$Li-{\it d} relative motion energy
at which the transition probability
reaches a maximum are sensitive to the choice of
the optical potential.
A standard set of the potential yields the branching
ratio of order 10$^{-4}$.

This work was supported by a Grant-in Aid for Scientific Research
(No. 05243102 and No. 06640381) of the Ministry of Education,
Science and Culture(Japan).




\narrowtext
\begin{table}
\caption{Dependence of the branching ratio of the $^{11}$Li $\beta$
decay into $^9$Li and {\it d} on a set of the
parameters, {\it a} and $r_0$ (in fm ), of the optical potential.
The hybrid model wave function is used.
}
\begin{tabular}{c|ccc}
   & a=0.5 & a=0.6 & a=0.7 \\ \hline
r$_0$=1.1&1.4$\times10^{-4}$ &2.2$\times10^{-4}$
         &4.7$\times10^{-4}$ \\
r$_0$=1.2&5.2$\times10^{-4}$ &0.5$\times10^{-4}$
         &1.0$\times10^{-6}$ \\
r$_0$=1.3&1.1$\times10^{-6}$ &5.2$\times10^{-6}$
         &9.5$\times10^{-6}$ \\
\end{tabular}
\label{Tab.1}
\end{table}

\begin{figure}
\caption{The contribution of the initial and final wave functions to
the GT matrix element.
Absolute values of $R$$g_E(R)$ for $E=0.5$ MeV and of
$R \psi_{eff}(R)$ and
the integral $I(R)$ are drawn.
See text for the definition of $I(R)$.
}
\label{fig1}
\end{figure}

\begin{figure}
\caption{
The transition probability  per time and energy units
as a function of the relative
motion energy {\it E}
for different sets of the optical potential parameters.
The values of $a$ and $r_0$ are given in units of fm.
The hybrid model wave function is used.
}
\label{fig2}
\end{figure}
\end{document}